\documentclass[aps,twocolumn,nofootinbib,showpacs]{revtex4}
\usepackage{bm,amssymb,graphicx}
\newcommand{\Qcommu}[2]{[#1,#2]}

\begin{document}
\bibliographystyle{apsrev}

\title{Gravity as a Yang-Mills Theory Based on the Lorentz Group}

\author{Hans Christian \"Ottinger}
\email[]{hco@mat.ethz.ch}
\homepage[]{www.polyphys.mat.ethz.ch}
\affiliation{ETH Z\"urich, Department of Materials, HCP F 47.2,
CH-8093 Z\"urich, Switzerland}

\date{\today}

\begin{abstract}
The Yang-Mills theory associated with the restricted Lorentz group is revisited as a candidate for a theory of gravity. This is a natural idea because the principle of equivalence of gravitation and inertia suggests to introduce locally inertial coordinate systems with the gauge freedom of Lorentz transformations. Compared to previous implementations of the idea, we use a generalized expression for the vector potential of the gauge theory in terms of the metric, which involves a coupling constant. One can verify that, in the limit of small coupling constant, all the classical predictions of general relativity are reproduced. For the resulting higher-order field equations, the nature of the singularity associated with black holes changes. The proposed theory is based on a dimensionless action for coupling the Yang-Mills field to matter.
\end{abstract}



\maketitle

Einstein's general theory of relativity is a shining example of how physical theories should be: it combines magnificent mathematical elegance with impressive predictive power. Despite the enormous success of Einstein's theory in all experimental tests, a large number of alternatives to general relativity have been proposed in the literature.

Einstein's general theory of relativity seems to have only one single defect, which is $90$ years of resistance to quantization (of course, the present search for ``dark energy'' could be interpreted as another failure of the theory of gravity on the largest scales \cite{CapozzielloDeLau11}). This defect provides the motivation to revisit the Yang-Mills theory based on the restricted Lorentz group as a natural alternative to general relativity (see \cite{CapozzielloDeLau11} for a broad review of gauge theories for gravity). Such a theory highlights the principle of equivalence of gravitation and inertia, where locally inertial coordinate systems come with the gauge freedom of Lorentz transformations.

Only two years after the proposal of Yang-Mills theory \cite{YangMills54}, R.~Utiyama established a connection between the gravitational field and the Yang-Mills theory based on the Lorentz group \cite{Utiyama56}. That paper is at the origin of what is now known as gauge gravitation theory \cite{CapozzielloDeLau11,IvanenkoSar83}. The original proposal by Utiyama has been criticized as ``unnatural'' by C.\,N.~Yang (see footnote~5 of \cite{Yang74}) and Yang's allegedly more natural proposal \cite{Yang74} has itself been massively criticized in Chapter~19 of \cite{BlagojevicHehl}. In this letter, we propose a more general transformation between the metric of general relativity and the vector potential of Yang-Mills theory and a non-standard coupling of the Yang-Mills field to matter.

The key idea of this letter is to define a Lorentz covariant gauge theory based on the Lorentz group in terms of the action
\begin{equation}\label{action}
   I = \int \left( - \frac{1}{4} F_{a\mu\nu}  F^{a\mu\nu}
   + \frac{2 \pi G R}{c^4} \, {\cal L}_{\rm mat} \right) d^4x .
\end{equation}
The first term is the standard action associated with the free Yang-Mills field $F_{a\mu\nu}$, where $a$ labels the generators of the underlying Lie algebra and $\mu,\nu$ are space-time indices (we consider a Minkowski space where $x^0=ct$ is the product of the speed of light $c$ and time $t$, and $x^1,x^2,x^3$ are the spatial coordinates). In the same way as the Minkowski metric $\eta_{\mu\nu} = \eta^{\mu\nu}$ [with signature $(-,+,+,+)$] is used to lower or raise space-time indices, the Cartan-Killing metric $K_{ab}=K^{ab}$ [defined and evaluated below] is used to lower or raise Lie-algebra indices. As the action of the field turns out to be dimensionless, we cannot simply add the action of matter in terms of the Lagrangian ${\cal L}_{\rm mat}$. The required factor is expressed in terms of Newton's constant $G$, the speed of light $c$, and a curvature scalar $R$ (with dimensions of ${\rm length}^{-2}$). Specifically, we consider the Lagrangian of a point particle with mass $m$ and trajectory $x(t)$,
\begin{equation}\label{Lagrangianmat}
   {\cal L}_{\rm mat} =
   - m c \, \sqrt{- g_{\mu\nu}\big(x(t)\big) \frac{d x^\mu(t)}{dt} \frac{d x^\nu(t)}{dt}}
   \, \delta^3(\bm{x}(t)-\bm{x}) .
\end{equation}
The common origin of the Yang-Mills field $F_{a\mu\nu}$, the curvature scalar $R$ in Eq.~(\ref{action}), and the metric $g_{\mu\nu}$ in Eq.~(\ref{Lagrangianmat}) and some implications of the action (\ref{action}) remain to be elaborated. The terms ``curvature scalar'' and ``metric'' should be used with caution because they do not play exactly the same role as in general relativity. The determinant of the ``metric'' has no influence on the Minkowski space integral in Eq.~(\ref{action}); it rather defines the relation between the velocity and momentum four-vectors and could be interpreted as a modification of scalar mass into an anisotropic mass tensor. The ``curvature scalar'' is required only for a proper coupling of the actions of matter and field.

As the subsequent development strongly relies on the Lorentz group, we briefly summarize its most important properties. The Lorentz group consists of the real $4 \times 4$ matrices that leave the Minkowski metric invariant. Our focus is on its Lie algebra ${\rm so}(1,3)$, which actually characterizes the restricted Lorentz group (the connected component containing the identity element). This Lie algebra has six base vectors, three of which generate the Lorentz boosts in the coordinate directions and the other three generate rotations around the coordinate axes. It is convenient to switch back and forth between the labels $a=1, \ldots 6$ for all six generators and the pairs $(0,1)$, $(0,2)$, $(0,3)$ for the boosts and $(2,3)$, $(3,1)$, $(1,2)$ for the rotations according to Table~\ref{tabindexmatch}. With these index conventions, we can introduce the generators of the restricted Lorentz group as
\begin{equation}\label{Lorentzgenerators}
   {T^{a \kappa}}_\lambda = \eta^{\tilde{\kappa}\kappa} \, {\delta^{\tilde{\lambda}}}_\lambda
   - {\delta^{\tilde{\kappa}}}_\lambda \, \eta^{\tilde{\lambda}\kappa} ,
\end{equation}
so that ${T^a}_{\kappa\lambda} = \eta_{\kappa\kappa'} {T^{a \kappa'}}_\lambda$ and $T^{a \kappa\lambda} = \eta^{\lambda\lambda'} {T^{a \kappa}}_{\lambda'}$ are antisymmetric in $\kappa$ and $\lambda$ (and also in $\tilde{\kappa}$ and $\tilde{\lambda}$).

From the definition (\ref{Lorentzgenerators}), we can determine the structure constants $f^{ab}_c$ characterizing the commutators
\begin{equation}\label{Lorentzstructure}
   \Qcommu{T^a}{T^b} = f^{ab}_c \, T^c ,
\end{equation}
and the Cartan-Killing metric
\begin{equation}\label{LorentzCartanKilling}
   K^{ab} = \frac{1}{4} f^{ac}_d f^{bd}_c .
\end{equation}
These quantities are related to the properties of the traces of the generators according to
\begin{eqnarray}\label{Lorentztrace12}
   & {\rm tr}(T^a) = 0, \quad {\rm tr}(T^a T^b) = 2 K^{ab} , & \\
   & f^{abc} = {\rm tr}(T^a T^b T^c) = \frac{1}{2} {\rm tr}(\Qcommu{T^a}{T^b}T^c)
   = f^{ab}_d K^{dc} . &
\label{Lorentztrace3}
\end{eqnarray}
The explicit result $K^{ab} = - \eta^{\tilde{\kappa}_a \tilde{\kappa}_b} \, \eta^{\tilde{\lambda}_a \tilde{\lambda}_b}$ means that $K^{ab}$ is the diagonal $6 \times 6$ matrix with diagonal elements $(1,1,1,-1,-1,-1)$, which is its own inverse. The structure constants can be specified as follows: $f^{abc}$ is $1$ ($-1$) if $(a,b,c)$ is an even (odd) permutation of $(4,5,6)$, $(1,3,5)$, $(1,6,2)$ or $(2,4,3)$ and vanishes otherwise. These structure constants satisfy the Jacobi identity
\begin{equation}\label{Jacobiid}
   f^{sb}_a f^{cd}_s + f^{sc}_a f^{db}_s + f^{sd}_a f^{bc}_s = 0 .
\end{equation}

\begin{table}
  \centering
  \begin{tabular}{|c|c c c c c c|}
  \hline
  $a$ & $1$ & $2$ & $3$ & $4$ & $5$ & $6$ \\
  \hline
  $(\tilde{\kappa},\tilde{\lambda})$ & $(0,1)$ & $(0,2)$ & $(0,3)$ & $(2,3)$ & $(3,1)$ & $(1,2)$ \\
  \hline
  \end{tabular}
  \caption{Correspondence between label $a$ for the base vectors of the six-dimensional Lie algebra ${\rm so}(1,3)$ and ordered pairs of space-time indices.}\label{tabindexmatch}
\end{table}

We are now ready to develop the details of the Yang-Mills theory (\ref{action}) based on the restricted Lorentz group. The definition of the relevant variables is very similar to that of the Ashtekar variables proposed for a canonical approach to gravity \cite{Ashtekar86,Ashtekar87}. We begin by decomposing the metric tensor in the following form,
\begin{equation}\label{localinertialdecomp}
   g_{\mu\nu} = \eta_{\kappa\lambda} \, {b^\kappa}_\mu {b^\lambda}_\nu.
\end{equation}
The inverse $\bar{b} = b^{-1}$ allows us to decompose the inverse metric tensor in the same way, $\bar{g}^{\mu\nu} = \eta^{\kappa\lambda} \, \mbox{$\bar{b}^\mu$}_{\kappa} \mbox{$\bar{b}^\nu$}_{\lambda}$. As Lorentz transformations leave the Minkowski metric in Eq.~(\ref{localinertialdecomp}) invariant, we can easily identify the gauge transformations of $b$ and $\bar{b}$ that leave the metric tensor invariant: $\delta{b^\kappa}_\mu = - \tilde{g} \Lambda_a {T^{a \kappa}}_\lambda {b^\lambda}_\mu$ and
$\delta\mbox{$\bar{b}^\mu$}_{\kappa} = \tilde{g} \Lambda_a {T^{a \lambda}}_\kappa
\mbox{$\bar{b}^\mu$}_{\lambda}$, where $\tilde{g} \Lambda_a$ characterizes infinitesimal Lorentz transformations. If one has $g_{0j}=0$, then one can choose also ${b^0}_j={b^j}_0=0$ (by boosting), ${b^0}_0=\sqrt{-g_{00}}$, and the spatial block of ${b^\kappa}_\mu$ as the unique positive definite square root of the spatial block of $g_{\mu\nu}$ (by polar decomposition and rotation). We refer to this convenient possibility as the symmetric gauge (in view of the nonsymmetric role of the two indices of ${b^\kappa}_\mu$ in Eq.~(\ref{localinertialdecomp}), this possibility is remarkable).

We next wish to construct the four-vector potential $A_{a \nu}$ for the Yang-Mills field $F_{a\mu\nu}$ that should possess the gauge transformation behavior
\begin{equation}\label{gaugeA}
   \delta A_{a \nu} = \frac{\partial \Lambda_a}{\partial x^\nu}
   + \tilde{g} f^{bc}_a \, A_{b \nu} \, \Lambda_c .
\end{equation}
For this four-vector potential, which has the interpretation of a connection between the variables ${b^\kappa}_\mu$ at different positions, we propose the following explicit representation (assuming $a=(\tilde{\kappa},\tilde{\lambda})$ according to Table~\ref{tabindexmatch}),
\begin{eqnarray}
    A_{a \nu} &=& \frac{1}{2} \,
    \mbox{$\bar{b}^\mu$}_{\tilde{\kappa}} \left( \frac{\partial g_{\mu\nu}}{\partial x^{\mu'}}
    -\frac{\partial g_{\mu'\nu}}{\partial x^\mu} \right) \mbox{$\bar{b}^{\mu'}$}_{\tilde{\lambda}}
    \nonumber\\
    &+& \frac{1}{2 \tilde{g}} \, \frac{\partial {b^\kappa}_\mu}{\partial x^\nu}
    \left( \mbox{$\bar{b}^\mu$}_{\tilde{\kappa}} \, \eta_{\kappa\tilde{\lambda}}
    - \eta_{\tilde{\kappa}\kappa} \, \mbox{$\bar{b}^\mu$}_{\tilde{\lambda}} \right) .
\label{Arepresentation}
\end{eqnarray}
One usually assumes the value $\tilde{g}=1$. We nevertheless introduce the variable coupling constant $\tilde{g}$ to exploit the full potential of Yang-Mills theory. Note that the factor $\tilde{g}$ between the two different contributions in Eq.~(\ref{Arepresentation}) can not be absorbed in the normalization of $A_{a \nu}$. The value of $\tilde{g}$ has no influence on the weak-field approximation or on the limit of Newtonian gravity. However, the presence of $\tilde{g}$ makes the full theory significantly different from previous proposals of Yang-Mills gravity based on the Lorentz group \cite{Utiyama56,Yang74}.

Knowing the gauge transformation behavior (\ref{gaugeA}) of the vector potentials $A_{a \nu}$, we can now build a standard Yang-Mills theory based on the restricted Lorentz group (see, e.g., Sect.~15.2 of \cite{PeskinSchroeder}, Chap.~15 of \cite{WeinbergQFT2}, or \cite{hco229}). The field tensor is obtained from the four-vector potential as
\begin{equation}\label{Fdefinition}
   F_{a \mu\nu} = \frac{\partial A_{a \nu}}{\partial x^\mu}
   - \frac{\partial A_{a \mu}}{\partial x^\nu}
   + \tilde{g} f^{bc}_a A_{b \mu} A_{c \nu} .
\end{equation}
To complete the discussion of the meaning of the fundamental action (\ref{action}), we define the curvature scalar $R$ in terms of $\Delta^{a \mu\nu} = \mbox{$\bar{b}^\mu$}_{\kappa} \mbox{$\bar{b}^\nu$}_{\lambda} T^{a \kappa\lambda}$ and $F_{a \mu\nu}$,
\begin{equation}\label{Riccidefs}
   {R^\mu}_\nu = \Delta^{a \mu\sigma} F_{a \sigma\nu} , \qquad R = {R^\mu}_\mu .
\end{equation}
For $\tilde{g}=1$, it can be shown that ${R^\mu}_\nu$ actually is the Ricci tensor associated with the metric (\ref{localinertialdecomp}), so that $R$ is the usual curvature scalar. An additional power of the determinant of $g_{\mu\nu}$ could be introduced into the definition of the variable $R$ in the action (\ref{action}).

We begin the discussion of the implications of the fundamental action in Eqs.~(\ref{action}) and (\ref{Lagrangianmat}) with the equation of motion for a particle in a given field. From stationarity of the action with respect to variations of the particle trajectory we find
\begin{equation}\label{particlemotion}
   \frac{d^2 x^\mu}{d\tau^2} =
   - \Gamma^\mu_{\nu\nu'} \frac{d x^\nu}{d\tau} \frac{d x^{\nu'}}{d\tau}
   - \left( c^2 g^{\mu\nu} + \frac{d x^\mu}{d\tau} \frac{d x^\nu}{d\tau} \right)
   \frac{\partial \ln |R|}{\partial x^\nu} ,
\end{equation}
where $\Gamma^\mu_{\nu\nu'}$ is the Christoffel symbol associated with $g_{\mu\nu}$ and $\tau$ is the proper time of the particle, that is, $c d\tau/dt$ is given by the square root in Eq.~(\ref{Lagrangianmat}). Only in regions of constant curvature $R$ a particle follows a geodesic line. The spatial projector of the gradient of $\ln |R|$ acts like an additional force driving a particle away from regions of high curvature, which is still independent of mass and hence consistent with the (weak) principle of equivalence of gravitation and inertia. We assume geodesic motion also for a test particle in regions of zero curvature, although it might be preferable to argue that the infamous self-interactions lead to constant but nonzero curvature at the particle position. Large self-interactions might actually suppress the gradient of $\ln |R|$ entirely.

Variation of the pure Yang-Mills action in Eq.~(\ref{action}) with respect to $A_{a \nu}$ leads to the usual gauge invariant field equation
\begin{equation}\label{YMfieldeqpure}
   \frac{\partial F_a^{\mu\nu}}{\partial x^\mu}
   + \tilde{g} f^{bc}_a A_{b \mu} F_c^{\mu\nu} = 0 ,
\end{equation}
which is a second-order differential equation for $A_{a \nu}$ and a third-order differential equation for ${b^\kappa}_\mu$. However, as $A_{a \nu}$ is defined in terms of ${b^\kappa}_\mu$ and its derivatives, the more appropriate variation with respect to ${b^\kappa}_\mu$ implies a fourth-order differential equation for ${b^\kappa}_\mu$, in which the further differential operator
\begin{eqnarray}\label{onemoreop}
   {{\cal D}^a_{\nu\nu'}}^{\mu'} &=&
   \left( \frac{\partial g_{\rho\nu}}{\partial x^{\nu'}} {\delta^{\mu'}}_{\!\!\sigma}
   - \frac{\partial g_{\nu'\sigma}}{\partial x^\rho} {\delta^{\mu'}}_{\!\!\nu} \right)
   \Delta^{a \sigma\rho} \\ \nonumber
   && \hspace{-2.3em} + \, \frac{\partial}{\partial x^\rho} ( g_{\nu\nu'} {\delta^{\mu'}}_{\!\!\sigma}
   + g_{\sigma\nu'} {\delta^{\mu'}}_{\!\!\nu} ) \, \Delta^{a \sigma\rho}
   + \frac{1}{\tilde{g}} \frac{\partial}{\partial x^\nu}
   g_{\rho\nu'} \Delta^{a \mu'\rho}
\end{eqnarray}
acts on the left-hand side of Eq.~(\ref{YMfieldeqpure}), where $a$ and $\nu$ are to be summed over.

We now turn to the weak-field approximation and the limit of Newtonian gravity. For ${b^\kappa}_\mu = {\delta^\kappa}_\mu + (1/2) \mbox{$\hat{h}^\kappa$}_\mu$, we can calculate $F_{a \mu\nu}$ and $R$ to first order in $\mbox{$\hat{h}^\kappa$}_\mu$. Both quantities depend only on the symmetric part $h_{\mu\nu} = ( \hat{h}_{\mu\nu} + \hat{h}_{\nu\mu} ) / 2$ and both are independent of $\tilde{g}$. Stationarity of the action (\ref{action}) with respect to variations of $h_{\mu\nu}$ implies the field equation
\begin{equation}\label{weakfield}
   \Box R_{\mu\nu}^{(1)} - \frac{1}{2}
   \frac{\partial^2 R^{(1)}}{\partial x^\mu \partial x^\nu} =
   - \frac{4 \pi G}{c^4} \left( \eta_{\mu\nu} \Box -
   \frac{\partial^2 }{\partial x^\mu \partial x^\nu} \right) {\cal L}_{\rm mat} ,
\end{equation}
where $R_{\mu\nu}^{(1)}$ and $R^{(1)}$ are the linearized versions of the quantities defined in Eq.~(\ref{Riccidefs}), $\Box = \partial^2 / \partial x^\mu \partial x_\mu$, and ${\cal L}_{\rm mat}$ is to be evaluated with $\eta_{\mu\nu}$ instead of $g_{\mu\nu}$. For time-independent mass distributions $\rho$, we expect ${\cal L}_{\rm mat} = - \rho c^2$, which is consistent with Eq.~(\ref{Lagrangianmat}) for point particles. If we then define a dimensionless Newtonian potential $\phi$ through the Poisson equation $\Box \phi = 4 \pi G \rho/c^2$, Eq.~(\ref{weakfield}) can easily be solved in two steps,
\begin{equation}\label{weakfieldsol}
   R_{\mu\nu}^{(1)} = \eta_{\mu\nu} \Box \phi
   + 2 \frac{\partial^2 \phi}{\partial x^\mu \partial x^\nu} , \quad
   h_{\mu\nu} = 2 \eta_{\mu\nu} \phi .
\end{equation}
The result $h_{00}=-2\phi$ implies $\Gamma^\mu_{00} = \partial \phi / \partial x_\mu$ so that geodetic motion at small velocities reproduces Newton's theory of gravity.

Another important feature of general relativity is the prediction of gravitational waves, which have been detected in 2015. If we are interested in plane waves travelling in the $3$-direction, according to Section~10.2 of \cite{Weinberg}, we can look for a linearized solution of the form
$h_{\mu\nu} = h ( \delta_{\mu 2} \delta_{\nu 2} - \delta_{\mu 1} \delta_{\nu 1}) - \tilde{h} ( \delta_{\mu 1} \delta_{\nu 2} + \delta_{\mu 2} \delta_{\nu 1})$ where $h$ and $\tilde{h}$ depend only on $t$ and $x_3$. In the absence of matter, Eq.~(\ref{weakfield}) becomes
\begin{equation}\label{gravwavesolh}
   \Box\Box h = \Box\Box \tilde{h} = 0.
\end{equation}
These equations characterize the same transverse modes as obtained from general relativity.


The classical high-precision tests of Einstein's theory of gravity are based on the general form of the static isotropic metric produced by a point mass at the origin. For this special type of metric, we can afford to solve the third-order Yang-Mills equation (\ref{YMfieldeqpure}) in matter-free space (of course, the fourth-order equations obtained by acting with the differential operator (\ref{onemoreop}) on Eq.~(\ref{YMfieldeqpure}) are solved, too). The static isotropic metric is of the general form
\begin{equation}\label{genisometricg}
   g_{\mu\nu} = \left(
            \begin{array}{cc}
             - B & 0 \\
             0 &
             \delta_{mn} + (A-1) \, x_m x_n/r^2 \\
            \end{array}
         \right) ,
\end{equation}
where $A$ and $B$ are functions of $r=(x_1^2+x_2^2+x_3^2)^{1/2}$. For the symmetric gauge, the corresponding fields $A_{a \mu}$ introduced in Eq.~(\ref{Arepresentation}) are of the form
\begin{eqnarray}
   A_{a 0} &=& - Z (
                   \begin{array}{cccccc}
                     x_1 & x_2 & x_3 & 0 & 0 & 0 \\
                   \end{array}
                 ), \nonumber\\
   A_{a 1} &=& Y (
                   \begin{array}{cccccc}
                     0 & 0 & 0 & 0 & x_3 & -x_2 \\
                   \end{array}
                 ), \nonumber\\
   A_{a 2} &=& Y (
                   \begin{array}{cccccc}
                     0 & 0 & 0 & -x_3 & 0 & x_1 \\
                   \end{array}
                 ), \nonumber\\
   A_{a 3} &=& Y (
                   \begin{array}{cccccc}
                     0 & 0 & 0 & x_2 & -x_1 & 0 \\
                   \end{array}
                 ),
\label{genisometricA}
\end{eqnarray}
with
\begin{equation}\label{YZexpressions}
   Y = \frac{1}{\tilde{g} \, r^2} \left[ 1 - (1-\tilde{g}) \frac{\sqrt{A}}{2}
   - \frac{1+\tilde{g}}{2\sqrt{A}} \right] , \quad
   Z = \frac{1}{2} \frac{1}{\sqrt{AB}} \frac{B'}{r} .
\end{equation}
The field equations (\ref{YMfieldeqpure}) lead to the following differential equations for the functions $Y$ and $Z$ characterizing the vector potentials,
\begin{eqnarray}\label{YZcondition2}
   Y'' + 4 \mbox{$\frac{Y'}{r}$} + \tilde{g} \, (3 - \tilde{g} r^2 Y) Y^2
   + \tilde{g} \, (1 - \tilde{g} r^2 Y) Z^2 \! & = & \! 0 , \qquad \\
   Z'' + 4 \mbox{$\frac{Z'}{r}$} + 2 \tilde{g} \, (2 - \tilde{g} r^2 Y) \, Y Z \! & = & \! 0 .
\label{YZcondition1}
\end{eqnarray}

We first analyze our results on the basis of the standard Robertson expansions (see, e.g., Sect.~8.3 of \cite{Weinberg}),
\begin{equation}\label{Robertsonexpansion}
  A = 1 + 2 \gamma \mbox{$\frac{MG}{r}$} + \ldots , \quad
  B = 1 - 2 \mbox{$\frac{MG}{r}$} + 2 (\beta-\gamma) \mbox{$\frac{M^2G^2}{r^2}$} + \ldots ,
\end{equation}
from which we obtain the corresponding leading-order terms of $Y$, $Z$,
\begin{equation}\label{RobertsonexpansionYZ}
   Y = \gamma \frac{MG}{r^3} + \ldots , \quad
   Z = \frac{MG}{r^3} - (2\beta-\gamma-1) \frac{M^2G^2}{r^4} + \ldots .
\end{equation}
The $1/r$ term in the expansion of $B$ defines the mass $M$. The other leading order terms in Eq.~(\ref{Robertsonexpansion}) introduce the coefficients $\beta$ and $\gamma$ which, according to Einstein's theory of general relativity, are both equal to unity. These coefficients are of crucial importance because they determine the famous weak-field predictions of the theory. Whereas Eq.~(\ref{YZcondition2}) is trivially satisfied to the given order, Eq.~(\ref{YZcondition1}) results in the condition
\begin{equation}\label{betagammarestriction}
   4 \beta - ( 2 + \tilde{g} ) \gamma -2 = 0 .
\end{equation}
For $\tilde{g}=1$, this condition is not consistent with the experimentally confirmed results $\beta=\gamma=1$ of Einstein's theory. If we insist on reproducing Einstein's prediction for the precession of perihelia (requiring $2\gamma-\beta=1$), then we find the factor $(1+\gamma)/2 = (1-\tilde{g}/6)/(1-\tilde{g}/3)$ characterizing the deflection of light by the sun. For $\tilde{g}=1$, Einstein's prediction for the deflection becomes increased by the factor $1.25$, which is far beyond the tolerance level of modern tests. In the limit of small coupling constants $\tilde{g}$, however, our result (\ref{betagammarestriction}) becomes fully consistent with Einstein's theory.

\begin{figure}
\centerline{\includegraphics[width=7 cm]{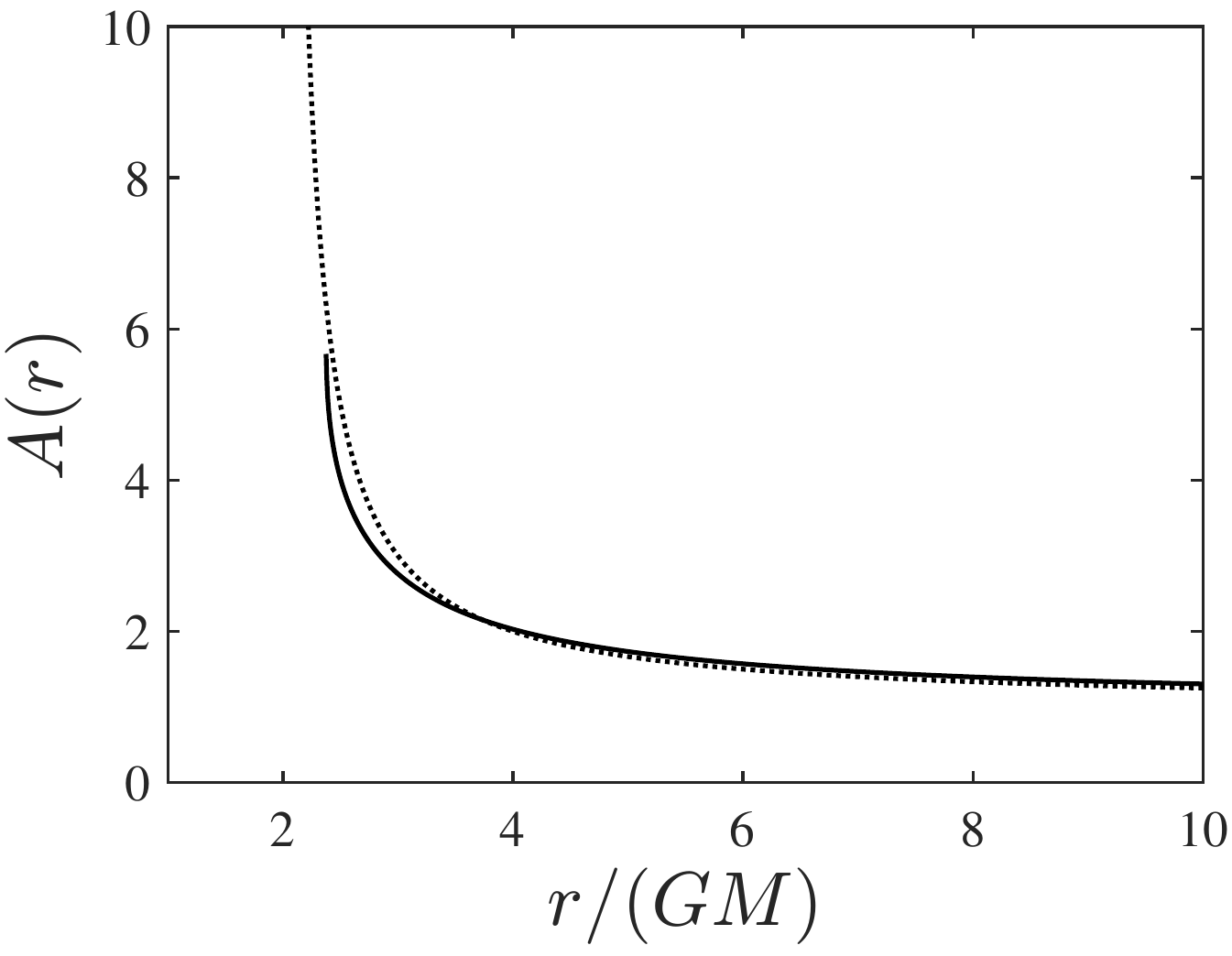}}
\caption[ ]{The function $A$ characterizing the spatial part of the isotropic metric (\ref{genisometricg}) obtained from the present Yang-Mills-type theory for $2\gamma-\beta=1$ and $\tilde{g}=0.7$ (continuous line) and for the Schwarzschild solution (dotted line).} \label{BRST_gravity_fig_Schwarzschild}
\end{figure}

Whereas the high-precision tests of general relativity rely on the asymptotic behavior of $A$ and $B$ for large $r$, we are also interested in seeing whether singularities arise around the Schwarzschild radius $r_{\rm s} = 2 G M$. We hence integrate Eqs.~(\ref{YZcondition1}) and (\ref{YZcondition2}) numerically (by substituting $z=1/r$ and translating Eq.~(\ref{RobertsonexpansionYZ}) into proper boundary conditions at $z=0$). The solution for $2\gamma-\beta=1$ and $\tilde{g}=0.7$ is shown in Figure~\ref{BRST_gravity_fig_Schwarzschild}. It is impressive that the present theory can reproduce the Schwarzschild solution over almost the entire range of $r$ values. However, more important is the observation that the nature of the singularity changes. Whereas the Schwarzschild solution diverges at $r_{\rm s}$, the solution of Eqs.~(\ref{YZcondition1}) and (\ref{YZcondition2}) ends when the slope diverges at finite values of $r$ and $A$. This change of the nature of the singularity is a consequence of dealing with a third-order rather than second-order field equation. This new type of singularity arises only for $\tilde{g}<1$ because then $1/\sqrt{A}$ is determined from $Y$ according to a quadratic rather than linear equation (\ref{YZexpressions}). In other words, the singularity is associated with a bifurcation.

The solution of Eqs.~(\ref{YZcondition1}) and (\ref{YZcondition2}) for $\tilde{g} \rightarrow 0$ admitting the proper limiting behavior $\beta, \gamma \rightarrow 1$ corresponds to the Schwarzschild metric of general relativity (see, e.g. \cite{Weinberg,MisnerThorneWheeler}). One easily verifies that this solution is given by $Y=Z=MG/r^3$. The expressions in Eq.~(\ref{YZexpressions}) imply $A=1$ and $B'/\sqrt{AB} = 2MG/r^2$. The latter relationship implies $2\beta-\gamma=1$ and actually holds rigorously for the Schwarzschild solution (for which one moreover has $A B = 1$, so that $B$ and $A$ can be found). For the present theory, the trivial result $A=1$ calls for a more careful consideration of the limit $\tilde{g} \rightarrow 0$. For small $\tilde{g}$ we find the leading-order behavior
\begin{equation}\label{limitresultA}
   \frac{1}{\sqrt{A}} = 1 + \tilde{g} \left( \sqrt{1 - \frac{r_0}{r}} - 1 \right) ,
\end{equation}
with $r_0=2MG/\tilde{g}$. For $\tilde{g}=1$ this result would actually coincide with the Schwarzschild solution, but only small values of $\tilde{g}$ are allowed in Eq.~(\ref{limitresultA}) (the values should not be too small because then $r_0$ becomes large; for all $\tilde{g}<1$, we obtain the same qualitative behavior). The corresponding result for $B$ is
\begin{equation}\label{limitresultB}
   \sqrt{B} = 1 - (1+\tilde{g}) \frac{MG}{r}
   + \frac{2}{3} \tilde{g} \frac{MG}{r_0} \left( 1 - \sqrt{1 - \frac{r_0}{r}}^3 \right) .
\end{equation}
The approximate black-hole solution (\ref{limitresultA}), (\ref{limitresultB}) leads to $\beta=\gamma=1$. For $\tilde{g}=2/3$, the singularity occurs at $r_0=3MG$.

In conclusion, we have proposed a generalized connection variable for interpreting the Yang-Mills theory based on the Lorentz group as a theory of gravity. This variable introduces an adjustable coupling constant into the Yang-Mills theory. For small coupling constants, the weak-field predictions of Einstein's general theory of relativity are reproduced. The general form of the static isotropic metric suggests that the singularity associated with black holes is shifted from the metric to its derivative. The coupling of the Yang-Mills field to matter is achieved by the action (\ref{action}). At least in matter-free space, quantization of the proposed theory can be achieved with the standard tools for Yang-Mills theories \cite{PeskinSchroeder,WeinbergQFT2,hco229}, allowing us to unify the theory of all interactions under the Yang-Mills umbrella.

Contrary to the effective field theories for the electroweak and strong interactions with an arbitrary cutoff that requires renormalization, the Yang-Mills theory for gravity should be regularized by a physical mechanism on the Planck scale. As suggested in \cite{hcoqft}, this can be achieved very robustly in the setting or irreversible thermodynamics, where the regularization is provided by dissipative smearing.


\end{document}